\documentclass[preprint,showpacs,preprintnumbers,amsmath,amssymb, showkeys]{revtex4}
\usepackage{graphicx}
\usepackage{dcolumn}
\usepackage{bm}
\usepackage{hyperref}

\begin{document}
	
	\title{Toward ternary fission accompanied by the ${}^{8}$Be nucleus\footnote[1]{To be published in Proceedings of  IX International Symposium on Exotic Nuclei (EXON-2018) 10 - 15 September, 2018, Petrozavodsk, Russia.}}
	
\author{D.A. Artemenkov}
\affiliation{Joint Institute for Nuclear Research, 141980, Dubna, Russia} 
\author{V. Bradnova}
\affiliation{Joint Institute for Nuclear Research, 141980, Dubna, Russia} 
\author{M.V. Gustova}
\affiliation{Joint Institute for Nuclear Research, 141980, Dubna, Russia} 
\author{O.M. Ivanov}
\affiliation{Joint Institute for Nuclear Research, 141980, Dubna, Russia} 
\author{R.R. Kattabekov}
\affiliation{Joint Institute for Nuclear Research, 141980, Dubna, Russia} 
\author{K.Z. Mamatkulov}
\affiliation{Joint Institute for Nuclear Research, 141980, Dubna, Russia} 
\author{V.V. Rusakova}
\affiliation{Joint Institute for Nuclear Research, 141980, Dubna, Russia} 
\author{A.V. Sabelnikov}
\affiliation{Joint Institute for Nuclear Research, 141980, Dubna, Russia} 
\author{ A.A. Zaitsev}
\affiliation{Joint Institute for Nuclear Research, 141980, Dubna, Russia} 
\author{P.I. Zarubin}
\email{zarubin@lhe.jinr.ru}
\affiliation{Joint Institute for Nuclear Research, 141980, Dubna, Russia} 
\author{I.G. Zarubina}
\affiliation{Joint Institute for Nuclear Research, 141980, Dubna, Russia}

\author{I. Ambro\u{z}ova}
\affiliation{Nuclear Physics Institute, Czech Academy of Sciences, 250 68 \u{R}e\u{z}, Czech Republic} 
\author{M. Kakona}
\affiliation{Nuclear Physics Institute, Czech Academy of Sciences, 250 68 \u{R}e\u{z}, Czech Republic} 
\author{M. Lu\u{z}ova}
\affiliation{Nuclear Physics Institute, Czech Academy of Sciences, 250 68 \u{R}e\u{z}, Czech Republic} 
\author{O. Ploc}
\affiliation{Nuclear Physics Institute, Czech Academy of Sciences, 250 68 \u{R}e\u{z}, Czech Republic} 
\author{K. Turek}
\affiliation{Nuclear Physics Institute, Czech Academy of Sciences, 250 68 \u{R}e\u{z}, Czech Republic} 

\author{E. Firu}
\affiliation{Institute of Space Science, 077125 Magurele, Romania} 
\author{M. Haiduc}
\affiliation{Institute of Space Science, 077125 Magurele, Romania} 
\author{A. Neagu}
\affiliation{Institute of Space Science, 077125 Magurele, Romania} 

\author{R. Stanoeva}
\affiliation{South-Western University, 2700 Blagoevgrad, Bulgaria} 

\pacs{21.60.Gx, 25.75.-q, 29.40.Rg}
\keywords{ternary fission, nuclear track emulsion, uranium, thermal neutrons, unstable nucleus}

\begin{abstract}
\indent Experiments in preparation for search for uranium  ternary fission by means of nuclear track emulsion are summarized. The study will be focused on the possible involvement of the unstable nucleus ${}^{8}$Be in the suggested scenario of the collinear tri-partition in the fission.

\end{abstract}

\maketitle 

\section*{Introduction}

\indent Implantation of uranium compounds into nuclear track emulsion (NTE) allow one to expand experimental means of nuclear fission studies. It is worth recalling that for the first time ternary and quaternary fission induced by neutrons was observed in such a way \cite{1}. For quite a while the ternary fission of the ${}^{252}$Cf isotope introduced into NTE has been explored \cite{2,3}. Owing its origin to NTE the physics of the ternary fission can serve as one of the drivers of interest in this classical technique. Not providing timing and energy analysis NTE preserves its observational mission. Due to full sensitivity to charged fragments in the full solid angle at record resolution of 0.5 $\mu$m this technique provides spatial patterns of few-prong nuclear fission events not unavailable in any other approach. Being measured lengths and thicknesses of tracks allow one to classify last ones as produced by as $\alpha$-particle, heavier or lighter fragment.\par 
Search for the collinear cluster tri-partition (CCT) of fissionable nuclei is among current challenges (reviewed in \cite{4}). Such a process is assumed to proceed through sequential binary fissions via formation of an intermediate state composed of two resulting fragments. Such a state is considered existing long enough with the respect to the fission time scale to be considered as a kind of a nuclear molecule. Decaying sequentially it could lead to alignment emission directions of the three fragment along the common axis. Therefore, there are obvious difficulties in separating pairs of fragments moving in the same direction. Orientation toward the ternary fission involving sufficiently long-lived isotope ${}^{8}$Be allows one to overcome this difficulty. ${}^{8}$Be emission in spontaneous decays ${}^{252}$Cf established recently \cite{5} supports this idea. It is worth noting that the ${}^{8}$Be emission mimics 2$\alpha$-particle radioactivity. In general, unbound configurations of lightest nuclei ($\alpha$,$t$) originated from decays of light nuclei exited above relevant thresholds aren’t excluded in the ternary fission.\par
\begin{figure}[t]
	\includegraphics[width=4in]{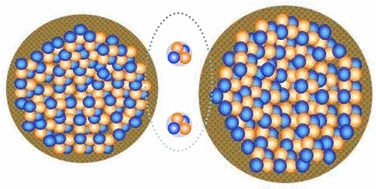}
	\caption{\label{Fig:1} Scenario of ternary fission of heavy nucleus involving ${}^{8}$Be.}
	\label{fig:Fig.1}
\end{figure}

The unstable ${}^{8}$Be nucleus is considered as a loose bond of $\alpha$-particles whose centers are a separated by a distance of about the $\alpha$-particle size. So, it would be too little to refer it to exotic nuclei. Due to its  size over deformation axis comparable to a heavy nucleus one ${}^{8}$Be can be considered as an important participant among heavier fission fragments (Figure \ref{fig:Fig.1}). Besides, ${}^{8}$Be could serve as a temporary covalent bond in an emerging ensemble. A ${}^{8}$Be accompanied fission  can proceed like 4-body instantaneous decay or sequential one when ${}^{8}$Be is kept by one of heavier fragments while the other one drifts away. The second option leads to the CCT pattern. Both scenarios seem intriguing and their interplay is possible. Thus, experimental examination of the pattern of the ternary fission involving ${}^{8}$Be is an inspiring goal.\par

Regarding the current status of the NTE technique the following can be noted. Being developed more than half a century ago it remains a status of a universal and cost-efficient detector. BR-2 type NTE with an unsurpassed observation beginning from fission fragments up to relativistic particles. In the last decade, the NTE technique is actively applied in the BECQUEREL experiment \cite{6} at the JINR Nuclotron allowing studying of light nuclei including radioactive ones nuclei at their relativistic dissociation \cite{7}. Unstable nuclei ${}^{8}$Be and ${}^{9}$B were identified by invariant masses of their decay products in the cases of the isotopes ${}^{10}$B and ${}^{10,11}$C \cite{8}.\par 
Meaning of the last fact is as follows. As is known, nucleosynthesis chains involving ${}^{8}$Be and ${}^{9}$B are suppressed due to an absence of the bound ground states. Nevertheless, this circumstance does not prevent the substantial structural contribution of ${}^{8}$Be and ${}^{9}$B. Therefore, it cannot be ruled out that they can serve as transient states persisting in the bound nuclei. The obtained experience was applied recently to identify for relativistic decays of the Hoyle-state \cite{9}. 
At the time, production of NTE layers in Moscow that had lasted for forty years stopped more than decade ago. The interest in further application of NTE stimulated its reproduction at the workshop MIKRON, which is a part of the Slavich Company (Pereslavl-Zalessky) \cite{10}. At present, NTE samples are produced by casting layers from 50 to 200 $\mu$m onto a glass substrate. The testing of the new NTE via irradiation with relativistic particles in above mentioned studies proved its similarity to NTE BR-2.\par 
Next, recent applications the reproduced NTE are summarized as essential ingredients in the suggested study (more detailed in \cite{11} and references herein). First of them is ability to reconstruct the ground ${}^{8}$Be$_{0^+}$ and first exited ${}^{8}$Be$_{2^+}$ states. The second one is sufficiency of angular resolution in direction determination of heavy ions relevant energy. Third one is preparation of NTE samples enriched with uranium. Last (and not least) is possibility of automatic counting of thermal neutron induced reactions in NTE.\par

\section*{Reconstruction of ${}^{8}$Be$_{2^+}$}

\begin{figure}[t]
	\includegraphics[width=4in]{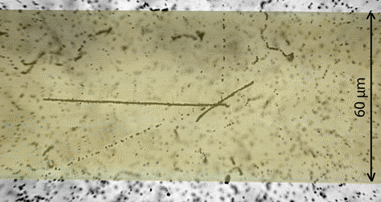}
	\caption{\label{Fig:2} Mosaic macro photograph of ``hammer-like" decay of the ${}^{8}$He nucleus stopped in the nuclear track emulsion (horizontal track). The decay results in a pair of relativistic electrons (dotted tracks) and a pair of $\alpha$ particles (oppositely directed short tracks). The decay image is superposed on a macro photograph of a human hair 60 $\mu$m thick to illustrate the spatial resolution.}
	\label{fig:Fig.2}
\end{figure}

\indent NTE were exposed to ${}^{8}$He nuclei with an energy of 60 MeV at the fragment separator ACCULINNA of the Flerov Laboratory of Nuclear Reactions (JINR). Figure \ref{fig:Fig.2} shows example of ``hammer-like" decay of ${}^{8}$He nucleus stopped in NTE as typical among about two thousand observed ones in this study. The search for $\beta$ decays of ${}^{8}$He nuclei was concentrated on the search for ``hammers". Often, gaps were observed between stopping points of primary tracks and ``hammer-like" decays in the so-called ``broken" events. ``Broken" events were assumed to take place owing to drift of the produced ${}^{8}$He atoms.\par

The directions of ${}^{8}$He arrival, the stopping points of their nuclei, the vertices of their decay, and the $\alpha$ particle stopping points were detected for 136 ``whole" and 142 ``broken" events. In ``broken" events, the decay points were determined by extrapolation of electron tracks. The distribution of opening angles for $\alpha$-particle pairs has an average value of (164.9 $\pm$ 0.7)$^\circ$. Such a ``hammer" kink connected with momentum carried away by an $e\nu$ pair is illustrating angular resolution.\par
\begin{figure}[t]
	\includegraphics[width=3in]{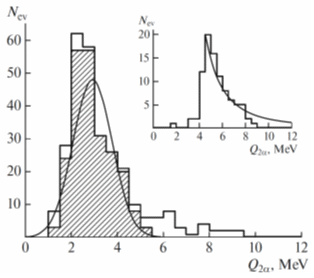}
	\caption{\label{Fig:3} Invariant mass $Q_{2\alpha}$ distribution for 278 pairs of $\alpha$ particles: shaded histogram corresponds $\alpha$ particles with ranges shorter than 12.5 $\mu$m; line corresponds to the Gaussian approximation; the inset shows the $Q_{2\alpha}$ distribution of 98 additional pairs with α-particle ranges above 12.5 $\mu$m.}
	\label{fig:Fig.3}
\end{figure}

The matching of an $\alpha$-particle range to its energy is obtained via spline interpolation of calculations using the program SRIM. The energies and opening angles of $\alpha$-particles yield the 2$\alpha$ invariant mass distribution $Q_{2\alpha}$. The variable $Q$ is determined as the difference between the invariant mass of the final system $M^*$ and the mass of the primary nucleus $M$. $M^*$ is determined as the sum of products of 4-momenta $P_{i,j}$, i.e., $M^{*2}$ = ($\sum{P_{i}}{P_{j}}$)$^2$. The $Q_{2\alpha}$ distribution (Figure \ref{fig:Fig.3}) on the whole corresponds to the decay from the first excited state of ${}^{8}$Be$_{2^+}$. \par
	
For events in which the ranges of both α particles are shorter than 12.5 $\mu$m and the opening angles are larger than 145$^\circ$, the average value of $Q_{2\alpha}$ is equal to (2.9 $\pm$ 0.1) MeV with the RMS equal to 0.85 MeV, which corresponds to ${}^{8}$Be$_{2^+}$. At the same time, this distribution has an extended ``tail" which is not described by a Gaussian. The inset in Figure \ref{fig:Fig.3} shows the $Q_{2\alpha}$ distribution corresponding to both ranges longer than 12.5 $\mu$m. It is possible that its shape reflects the spatial structure of the ${}^{8}$Be$_{2^+}$ state.\par

\section*{Reconstruction of ${}^{8}$Be$_{0^+}$}
\indent NTE exposed to 14.1 MeV neutrons allows one to study the ensembles of triples of $\alpha$-particles produced in disintegrations of ${}^{12}$C nuclei of NTE composition. Exposure of NTE to 14.1 MeV neutrons was performed on one of devices DVIN of an applied destination. In 400 selected 3$\alpha$ disintegrations measurements of angles relative to plane of a NTE layer and its surface as well as their lengths were performed for all $\alpha$-particle tracks. \par
Distribution over ranges of $\alpha$-particles $L_{\alpha}$ has an average value $\langle L_{\alpha} \rangle$ = (5.8 $\pm$ 0.2) $\mu$m at RMS (3.3 $\pm$ 0.1) $\mu$m. This distribution has an asymmetric shape described by the Landau distribution. Directly associated with it the distribution over energy of $\alpha$-particles $E_{\alpha}$ defined by ranges $L_{\alpha}$ in the SRIM model  has an average value $\langle E_{\alpha} \rangle$ = (1.86 $\pm$ 0.05) MeV and RMS (0.85 $\pm$ 0.03) MeV.\par

\begin{figure}[t]
	\includegraphics[width=5in]{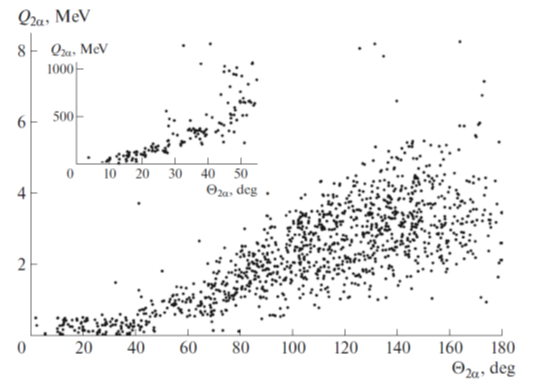}
	\caption{\label{Fig:4} Energy $Q_{2\alpha}$ and opening angle $\Theta_{2\alpha}$ correlations in $\alpha$-particle pairs produced in ${}^{12}$C splitting by 14.1 MeV neutrons.}
	\label{fig:Fig.4}
\end{figure}

\begin{figure}[t]
	\includegraphics[width=4in]{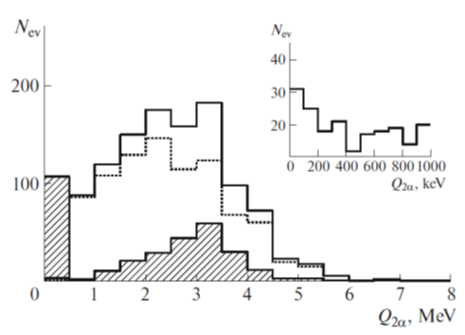}
	\caption{\label{Fig:5} Distribution of $\alpha$-particle pairs produced in ${}^{12}$C splitting by 14.1 MeV neutrons over invariant mass $Q_{2\alpha}$: (solid line) all $\alpha$ pairs, (shaded) selected ${}^{8}$Be$_{0^+}$, and (dashed line) their difference.}
	\label{fig:Fig.5}
\end{figure}

\begin{figure}[t]
	\includegraphics[width=4in]{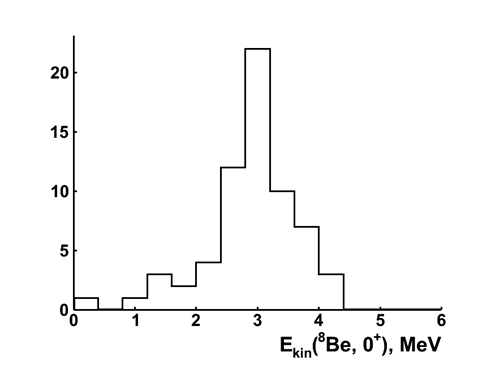}
	\caption{\label{Fig:6} Distribution ${}^{8}$Be$_{0^+}$ of $\alpha$-particle pairs produced in ${}^{12}$C splitting by 14.1 MeV neutrons over kinetic energy $E_{kin}$.}
	\label{fig:Fig.6}
\end{figure}

Determination of angles and energy values versus ranges allows one to determine the invariant mass $Q_{2\alpha}$ of pairs. Correlation over $Q_{2\alpha}$ and opening angles $\Theta_{2\alpha}$ in $\alpha$-particle pairs reveals features of the ${}^{8}$Be nucleus. The region of large opening angles $\Theta_{2\alpha}$ $>$ 90$^\circ$ is corresponding to $Q_{2\alpha}$ of ${}^{8}$Be$_{2^+}$, while $\Theta_{2\alpha}$ $<$ 30$^\circ$ − ${}^{8}$Be$_{0^+}$. Distribution over $Q_{2\alpha}$ points to these states (Figure \ref{fig:Fig.4}). Its right side meets the shape expected from the decay through ${}^{8}$Be$_{2^+}$. Condition $Q_{2\alpha}$ $<$ 200 keV has allowed to allocate 56 decays ${}^{8}$Be$_{0^+}$ (Figure \ref{fig:Fig.5}). The ${}^{8}$Be$_{0^+}$ selection reveals in the correlated peak relevant to the ${}^{8}$Be$_{2^+}$ excited state.\par
Reconstruction of the decaying ${}^{8}$Be$_{0^+}$ allows one to derive distribution over their values of kinetic energy $E_{kin}$ (Figure \ref{fig:Fig.6}). It has an average value $\langle E_{kin} \rangle$ = (3.0 $\pm$ 0.1) MeV and RMS (0.75 $\pm$ 0.07) MeV. Since ${}^{8}$Be$_{0^+}$ is reconstructed at lowest energy release it has to be reconstructed in the ternary fission when extra repulsion is provided by highly charged fragments.\par

\section*{Slow heavy ions in NTE}
\begin{figure}[t]
	\includegraphics[width=5in]{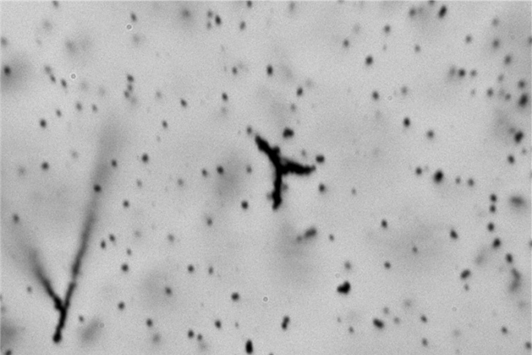}
	\caption{\label{Fig:7} Macro photograph of the first tri-partition event found in this study.}
	\label{fig:Fig.7}
\end{figure}

\indent Surface irradiations of NTE samples at the Department of Radiation Dosimetry of the Nuclear Physics Institute were performed at first with manual movement of the ${}^{252}$Cf source. Then a specially developed device was applied; the source was moved over the surface of this device automatically according to a convenient space and time pattern. When the NTE surface irradiated by the Cf source was examined, planar trios consisting of pairs of fragments and long-range $\alpha$ particles and trios of fragments were found (see Figure \ref{fig:Fig.7}). \par

\begin{figure}[t]
	\includegraphics[width=4in]{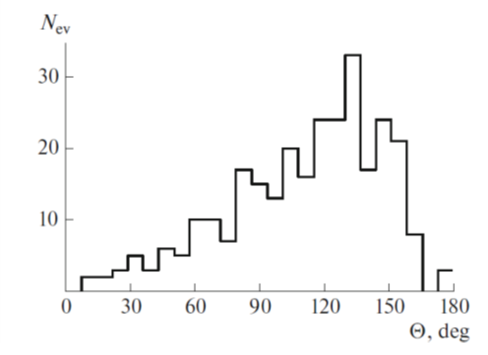}
	\caption{\label{Fig:8} Distributions of ${}^{252}$Cf tri-partition events over opening angles of fragment pairs.}
	\label{fig:Fig.8}
\end{figure}

The fragment opening angles were also measured in these events (Figure \ref{fig:Fig.8}). Their distribution is characterized by an average value of (111 $\pm$ 2)$^\circ$ and an RMS of 36$^\circ$. It can be concluded that no candidates for collinear fission have been found yet, and their search should be continued. The ranges of all fragments were measured in 96 events of  tri-partition without $\alpha$ particles. The average energy of fission fragments is about 400 A keV.\par

\begin{figure}[t]
	\includegraphics[width=5in]{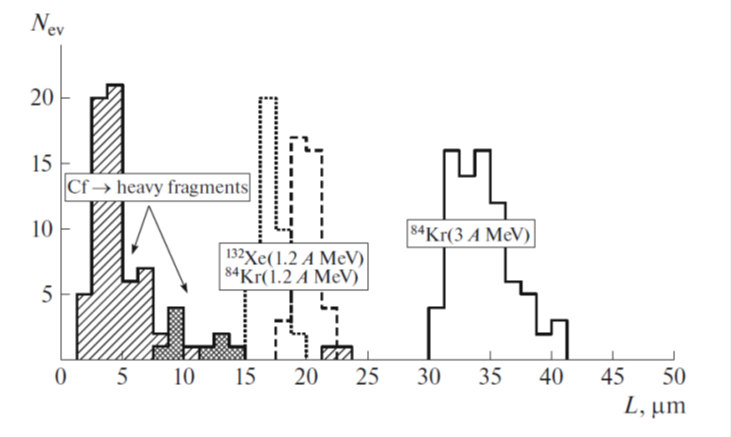}
	\caption{\label{Fig:9} ${}^{84}$Kr, ${}^{132}$Xe, and ${}^{86}$Kr ion range distributions in decays Cf $\to$ 3 fragments and Cf $\to$ 2 fragments + long-range $\alpha$.}
	\label{fig:Fig.9}
\end{figure}

Besides, NTE samples were exposed at the Flerov Laboratory of Nuclear Reactions (JINR) at the cyclotron IC-100 to 1.2 A MeV ${}^{86}$Kr$^{+17}$ and ${}^{124}$Xe$^{+26}$ ions and at the cyclotron U-400M to 3 $A$ MeV ${}^{84}$Kr ions. The irradiation was performed in vacuum irradiation chambers without a black paper using a red photo lamps as a light source. For better track observation, samples were installed at a large angle with respect to the beams. The sample irradiation density reached 10$^6$ tracks per cm$^2$ in several seconds. Figure \ref{fig:Fig.9} shows the range distribution for ions stopped in NTE without visible scattering. Then, the calibration of ion ranges in NTE needs to be extended much lower than 1 $A$.\par 

Reducing energy of the IC-100 beam is possible in two ways. The first and most simple one is  installation directly in front of a target aluminum film or Mylar desired thickness. Its advantage consists in immutability of the measured particle flux, and relatively fast replacement of degrader, but the drawback is that after degrader is producing complex flows over ion charges, angles and energy. The second method is carried out by installing a degrader immediately at the output of the accelerator at the beginning of a long transport channel. Its advantage lies in the fact that due to collimation in the channel ions fall on the target at normal, remaining mono-energy and mono-charge, and the lack of a long time of replacement of the degrader associated with the inlet of the atmosphere and its subsequent pumping, as well as a significant loss of the beam.\par
Recently, the idea of a smooth energy regulator was implemented. Both polymer and metal films (aluminum) can be used as a degrader in the regulator. When the frame degrader is rotated around the axis an effective path length of ions through the degrader varies according to the inverse cosine law. When the ion speed slows down, it is recharged. The recharge curve was calculated by the LISE program. This allowed to correctly predict the flux density of ions at the integrated current with the ``quick flap of the bundle". This makes it possible to move the calibration downwards in energy.\par

\section*{Exposure to thermal neutrons}
\indent NTE samples enriched in boron (boric acid and borax) were irradiated for 30 min at the thermal neutron channel of the IBR-2 reactor (JINR). The subsequent sample analysis yielded an intensity estimate of 5 $\times$ 10$^5$ neutrons per second. The samples were produced by casting boron-enriched NTE to a thickness of about 60 $\mu$m onto a 2 mm glass substrate. The application of glass resulted in activation of sodium contained in it, which presented a problem, although inevitable.\par
\begin{figure}[t]
	\includegraphics[width=4in]{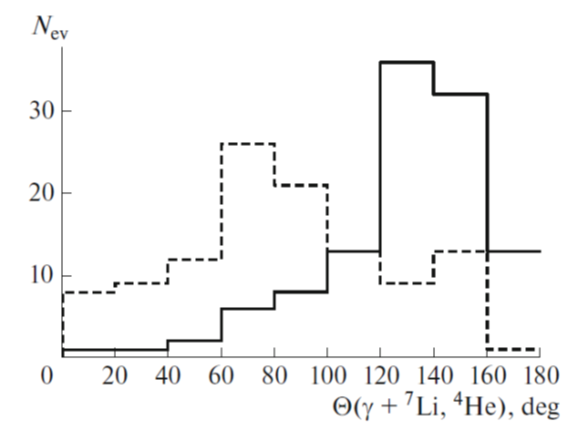}
	\caption{\label{Fig:10} Distribution over the angles $\Theta$($\gamma$ + ${}^{7}$Li,${}^{4}$He) between the calculated photon directions and the directions of (solid line) ${}^{7}$Li and (dashed line) ${}^{4}$He nuclei.}
	\label{fig:Fig.10}
\end{figure}

The presence of boron in NTE allows one to observe charged products of the $n_{th}$ + ${}^{10}$B $\to$ ${}^{11}$B$^*$ $\to$ ${}^{7}$Li$^*$ + ($\gamma$) + $\alpha$ reaction. The fragments take away energy of 2.8 MeV. A 478 keV photon is emitted with a probability of 93\% by the ${}^{7}$Li nucleus from its only excited state. The coordinate measurements of tracks in 112 ${}^{7}$Li + ${}^{4}$He events were performed. Their directions in pairs are no collinear and have an average angle $\Theta$(${}^{7}$Li + ${}^{4}$He) of (148 $\pm$ 14)$^\circ$ due to photon emission. The average energy $Q$(${}^{7}$Li + ${}^{4}$He) was (2.4 $\pm$ 0.2) MeV with RMS of 0.8 MeV in agreement with the energy carried away by the photon. The distribution of $\Theta$($\gamma$ + ${}^{7}$Li) between the photon emission angles calculated from the momentum conservation and the ${}^{7}$Li direction point to apparent anti-correlation (Figure \ref{fig:Fig.10}). It would be useful to calibrate in the reaction of $n_{th}$ + ${}^{6}$Li $\to$ $t$ + $\alpha$ in NTE enriched with the lithium borate. \par

\section*{Impregnation with Uranium}
\indent Recently, an impregnation with an aqueous solution of nitrate uranyl of NTE on a glass plate was mastered in the Flerov Laboratory. 20 samples of  NTE applied to one side of glass plates of 6 $\times$ 4.5 cm$^2$ were used. The procedure was conducted in a dark room at room temperature. Drying of samples was carried out under the same conditions. A lamp with a red light filter were used for lighting. It was established in advance that saturation of the samples with water is achieved in no more than 15 minutes.
A freshly prepared aqueous solution of UO$_2$(NO$_3$)$_2$ was poured into a Plexiglas cuvette with grooves for placing samples in an upright position. A thickness of a solution layer in a cuvette was about 20 mm. An initial concentration of a uranium solution determined by x-ray fluorescence analysis (XRF) was 600 $\mu$g/ml.\par
A soaking was conducted in 2 stages. In the first stage, 12 samples were soaked, and in the second stage, the remaining 8. The first 12 samples were pre-unpacked and then sequentially one after another installed in a ditch with a solution. The installation procedure took no more than 10 minutes. In the established position, the samples did not touch each other, the distance between the two adjacent samples was about 3 mm. 60 minutes after the installation of the last sample, the samples were extracted from the cuvette with the solution in the same order. The extraction procedure took no more than 10 minutes. During the extraction process, each sample was placed in a vertical position at the bottom edge on a sheet of filter paper to remove the residual solution, and then moved to a dry cell of Plexiglas with grooves to install the samples in a vertical position. Further drying of the samples took place in an upright position. In the established position the samples did not touch each other, the distance between two adjacent samples was about 3 mm. After extraction of all 12 samples from the solution, an aliquot was selected from the cuvette with the solution to determine the possible change in the concentration of uranium in the solution at the first stage.\par
At the second stage, the remaining 8 samples were immersed in the same solution. The mode of placing the samples in the solution, impregnation and extraction of samples from the solution was similar to the mode for 12 samples. After the extraction of all 8 samples from the solution, an aliquot was selected from the solution cell to determine the possible change in the uranium concentration in the solution at the second stage. The analysis of the uranium solution aliquots selected at the first and second stages by XRF showed a change in the concentration of uranium in the solution from 600 $\mu$g/ml to 500 $\mu$g/ml at the first stage and from 500 $\mu$g/ml to 410 $\mu$g/ml at the second stage. This gives reason to assume the presence of sorption properties of the emulsion with respect to uranium.\par

\begin{figure}[t]
	\includegraphics[width=5in]{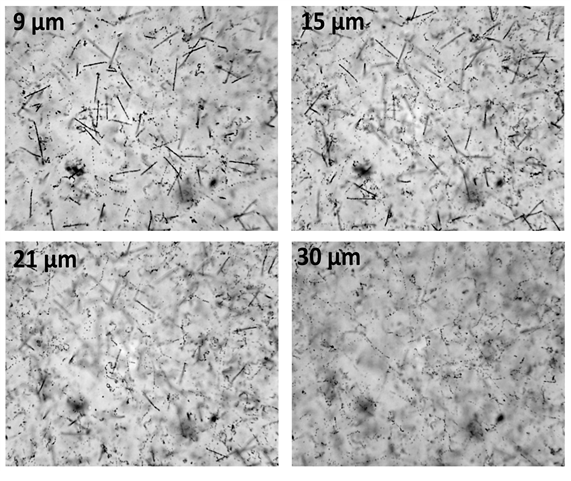}
	\caption{\label{Fig:11} Macro photographs of NTE soaked in solution of nitrate uranyl.}
	\label{fig:Fig.11}
\end{figure}

\begin{figure}[t]
	\includegraphics[width=4in]{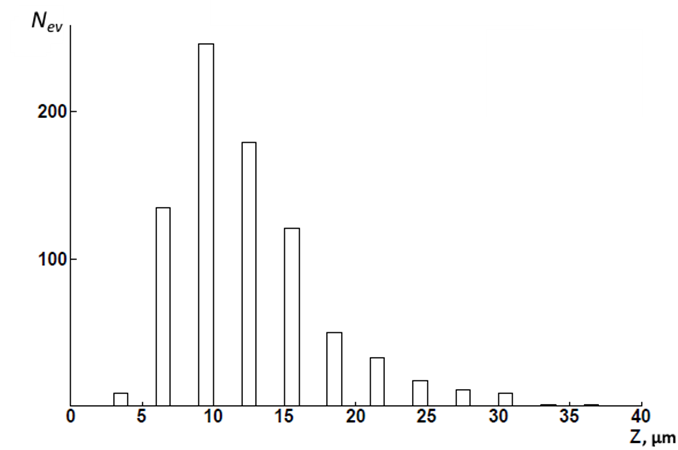}
	\caption{\label{Fig:12} Distribution of number of planar $\alpha$-tracks per 1 $\mu$m over immersion depth in NTE sample soaked in solution of uranyl nitrate.}
	\label{fig:Fig.12}
\end{figure}

The samples were developed in about two days. On macro photographs taken in one of the developed samples at immersion depths of 9, 15. 21 and 30 $\mu$m from the surface tracks of $\alpha$-particles are visible (Figure \ref{fig:Fig.11}). An observer can notice that their density decreases rapidly. Samples from both batches were scanned over an area of 16 points 1 cm apart. In each point were carried out the photographing depth of field with a step of 1 $\mu$m with a microscope objective $\times$60 and adapter for camera $\times$0.5.  Calculation of traces lying in the plane of the sample was carried out visually on the obtained images with a step in depth of 3 microns. It was found that in the sample from the second batch the number of $\alpha$ tracks is less by (10 $\pm$ 3) \% than in the samples from the first batch.  
The disappearance of tracks of $\alpha$-particles to a depth of no more than 40 $\mu$m was observed at all control points on the three studied samples. The total distribution of all found planar traces by depth (Figure \ref{fig:Fig.12}) gives an idea of this effect. The average value of the penetration depth of uranium is 15 $\mu$m with 6 $\mu$m RMS. It can be concluded that the absorption of uranium had proceeded in the surface layer of gelatin (in fact, protein medium).\par 

In the context of the proposed study, this effect indicates that the search for fission events should be almost planar, which is a fundamental simplification. In general terms, this observation is important for understanding the danger of uranium compounds to humans. At the same time, it allows us to offer methods of cleaning from contamination with gelatin.\par

\section*{Concluding remarks}
\indent The competitive character of NTE was proved recently in a series of measurements of short tracks of $\alpha$ particles and slow heavy ions. For these purposes microscopes KSM-1 manufactured by Carl Zeiss (Jena) about half of century and still functioning well are applied in Laboratory of High energy Physics, JINR. The possibility of $\alpha$ spectrometry was verified and the ${}^{8}$He atom drift effect was established in measurement of decays of ${}^{8}$He nuclei implanted in NTE. Correlations of $\alpha$ particle trios in ${}^{12}$C nucleus splitting by 14.1 MeV neutrons as well as ${}^{7}$Li and ${}^{4}$He nuclei produced in ${}^{10}$B break up by thermal neutrons in boron-enriched NTE were studied. In this series of experiments, the NTE resolution proved to be perfect as expected physical effects in invariant mass distributions of reaction product ensembles were clearly observed. NTE samples are calibrated using 1.2 and 3 $A$ MeV Kr and Xe ions. NTE surface exposures to a ${}^{252}$Cf source allowed to find and measure events containing fragment pairs and long range $\alpha$ particles, as well as fragment triples.\par
Thus, the major prerequisites are available to argue application of NTE technique in the state-of-art problems of the nuclear fission physics. Due to perfect observation the ${}^{8}$Be accompanied channel of the ternary fission can become the ``Golden key" for experimental verification of the CCT hypothesis. It is possible to be identifiable ${}^{8}$Be decays in NTE by measurements of decay $\alpha$-particle ranges and emission angles. Such events could be found among 4-prong ones. Smaller thickness and larger ranges of $\alpha$-particle ranges in them will allow to distinct them from quaternary fission cases. Observation of heavy fragments is a principal advantage NTE allowing studying angular correlations.\par 
Use of microscopes is last but not least aspect The use of modern computerized and automatic microscopes for the analysis of nuclear emulsion, containing both fission events and associated traces of radioactivity, allows one to address to tasks of the past at the state-of-art level that were considered exhausted due to the limited possibilities of direct human analysis. Being provided with relevant image recognition programs they could be suggested to finding and measuring short nuclear tracks with the most precise spatial resolution at an unprecedented statistics level. However, recognition of few-prong events in NTE has not yet been achieved with an automated microscope, especially, when real study complications are taken into account. At the present stage, automatic analysis can very effectively complement the human search for rare triple and fourfold fission events in monitor counting of binary fission and α-radioactivity tracks. Computerization of measurements on the microscope and automatic changing lenses to higher magnification when a relevant event is found can radically enhance researcher capabilities. Suitable microscopes are available on the market.\par 

\section*{Acknowledgments}
\indent Authors are grateful to Prof. S. N. Dmitriev for possibility to carry out described experiments in the Flerov Laboratory of JINR. Authors are especially indebted to Prof. Yuri Penionzhkevich (Flerov Laboratory, JINR) who attracted attention to possibility to use the NTE technique in the CCT problem. Saving of the NTE technique wouldn’t be possible without support of Prof. A. I. Malakhov  (Veksler \& Baldin Laboratory, JINR).

\end {document}